*Review Paper*

# Smart Traffic Systems: A Comprehensive Review of Recent Advancements, Technologies, and Challenges

**Arindom Chakraborty [1], Mehedi Hasan [1], Amzad Hossain [2,*], Meratun Junnut Anee [2]**

[1] Department of Electrical and Electronic Engineering, University of Science and Technology Chittagong, Chattogram 4202, Bangladesh.
[2] Department of Electrical and Computer Engineering, North South University, Dhaka-1229, Bangladesh.
- Corresponding Author: Amzad Hossain (e-mail: amzad.hossain01@northsouth.edu)

**Abstract:** With an ever-growing urban population, the need for transportation is increasing at an alarming rate. Thus, the massive increase in the number of vehicles is creating traffic congestion which creates various environmental, societal, and economic problems. To tackle traffic-related issues, several Smart Traffic Systems (STS) have been proposed and implemented. As a result, a comprehensive review of STS has become necessary. The main objective of this paper is to provide an overview and a thorough review of the existing STSs in terms of various technological approaches, traffic detection technologies using different sensors, various networking/communication tools, and their pros and cons. The paper also provides information on major STS services. In addition, challenges related to modern STS are identified. Therefore, the taxonomy of STSs provided in this paper will aid researchers, urban planners, and policymakers to recognize and install the best-suited STSs for their settings.

## 1. Introduction

Advances in science and technology have made our lives simpler, safer, and more secure. In the present days of modern science and technology, communication among various devices has gained quite a bit of momentum due to the rapid improvement of the internet, communication, and networking technologies. This technological advancement has led the world to an era of IoT or the Internet of Things [1]. Like many other innovations, IoT has paved the path to greater innovations such as the concept of a Smart City. The purpose of a smart city is to improve the lifestyle of its citizens by reducing operational costs, improving productivity, and effective management of its resources [2]. The smart city concept incorporates systematic surveillance and management of infrastructure, smart traffic and transportation systems [3], production and consumption of energy, healthcare [4,5], education, security, and many more sectors of city life.

With the rapid growth of the population in urban areas, the number of vehicles has increased by a couple of folds. Therefore, traffic management is a major concern in any modern metropolis. Megacities severely face traffic congestion issues, which waste valuable time, money, and fuel. In 2020, as per a poll from traffic allied market research by Inrix [6], citizens in Los Angeles and New York wasted 102 and 100 hours yearly just in traffic jams. Citizens of Moscow lose around 91 hours per year. On the other hand, people in London and Paris waste 74 hours and 69 hours per year, respectively [6]. According to a study conducted by the Harvard School of Public Health, traffic congestion had an impact of around $100 billion on the United States Economy in 2020 [7]. Similarly, the report published in 2019 by the Australian Infrastructure Audit claims cost due to traffic congestion in 2016 was $19 billion, which will rise to $39 billion in the year 2031 [8]. On the other hand, London and Berlin suffered $12.2 billion and $7.5 billion in damages due to traffic congestion in 2020 [9]. As the world’s population and urbanization increase, such costs will certainly be on the rise.

According to [10,11], traffic congestion increases the fuel consumption rate of vehicles, resulting in increased emissions of harmful gases like Carbon Monoxide (CO), Carbon Dioxide ($CO_2$), Volatile Organic Compounds (VOCs), Hydrocarbons (HCs), and Nitrogen Oxides ($NO_x$). Automobiles emit more than 5.53 million CO2 gases into the atmosphere each year, accounting for 16% of global emissions, according to 2011 data [12]. Similarly, motorized cars emit almost 72 % of $NO_2$ gas [13]. Many of the gases emitted by vehicles fall into the Greenhouse gas category, which depletes the Ozone layer and increases the overall average temperature of the world. According to National Oceanic and Atmospheric Administration (NOAA), Global

Climate Report-Annual 2020, the year 2020 was the second warmest year in 141-year records with a +0.98-degree Celsius departure in ocean surface temperature [14]. This sort of temperature difference has an adverse impact on agriculture, wildlife, and the water level of the oceans. Also, it creates an environmental and ecological imbalance throughout the world. Similarly, the gases emitted from vehicles are one of the main reasons for air pollution inflicted diseases. According to the United Nations Environment Program (UNEP) around 7 million premature deaths happen due to air pollution inflicted diseases such as bronchitis, pneumonia, etc. [15]. On the other hand, suspended particles alone are responsible for roughly more than 30,000 unanticipated deaths each year all over the world [16].

Since traffic congestion in megapolises imposes several major problems, an effective traffic management system is inevitable for the smooth and secure operation of these urban forests. As a result, the concept of Smart Traffic System (STS) has emerged. STS is the strategy of using modern technology to monitor, manage, and control traffic-related issues using the least possible resources. STS frameworks utilize sensors, communication modules, networking technology, and computational techniques to screen, coordinate, and control traffic-related activities. It also includes utilizing Global Positioning Systems (GPS) and combining it with sources from Wi-Fi, Bluetooth, and navigation tools to deliver up-to-the-minute positioning and traffic routing [17-19]. This advanced technology not only ensures efficient transportation but also promotes a safer and more convenient environment. In addition, smart technologies have the potential to revolutionize the way traffic is controlled and managed. One aspect of STS involves the placement of traffic lights, speed limit signs, and other road signals that work harmoniously to provide a consistent flow of traffic. These intelligent traffic control mechanisms make use of real-time data and analysis to optimize traffic patterns and prevent congestion [20-21]. Furthermore, these systems can seamlessly integrate with other emerging technologies, such as self-driving vehicle systems and emissions control methods, to further enhance traffic efficiency and reduce the environmental impact [22-24]. Another methodology used in smart traffic systems is the utilization of camera systems and radar technologies. These systems capture and analyze data about the traffic flow, enabling authorities to make informed decisions regarding traffic control, route optimization, signal timing, and vehicle detection [25-26]. The concept of modern STS is not limited to only these. Rather it includes traffic congestion detection, traffic prediction, accident detection, traffic speed adjustment, congestion avoidance, and many more features.

Due to various functionalities and features, several STS have been developed by researchers over the course of time. Due to the widespread research activities happening in the STS domain, it has become quite necessary to have a comprehensive review. Such review work will provide novice researchers, industry professionals, and policymakers with an overall overview of state-of-the-art works. Moreover, a comprehensive study will provide the basic knowledge to choose computational processes, sensors, communication protocols, and networking technologies as per the system requirements and environment.

This research aims to provide a bird's eye view of STS by conducting a comprehensive review of different aspects. The main contributions of this paper are:

- To provide an overview of STS
- To provide a comprehensive study on different approaches to STS
- To provide an overall study on the sensors utilized in various STS
- To provide a summary of communication and networking technologies
- To provide a summary of the suitability of STS
- To provide an overview of different STS features

The overall structure of this paper is provided in Figure 1. After providing introductory information and motivation for this paper in section 1, detailed research methodologies are presented in section 2. Section 3 provides an overall overview of STS. Section 4 discusses the major approaches used to design STS in various research works. Next, section 5 provides a detailed description of the different types of sensors utilized in STSs. After that, networking technologies are highlighted in section 6. A qualitative analysis of different STSs and their features is provided in section 7. In section 8, major challenges in future STSs are highlighted. Finally, concluding statements and future scope of work in STS field are provided in section 9.

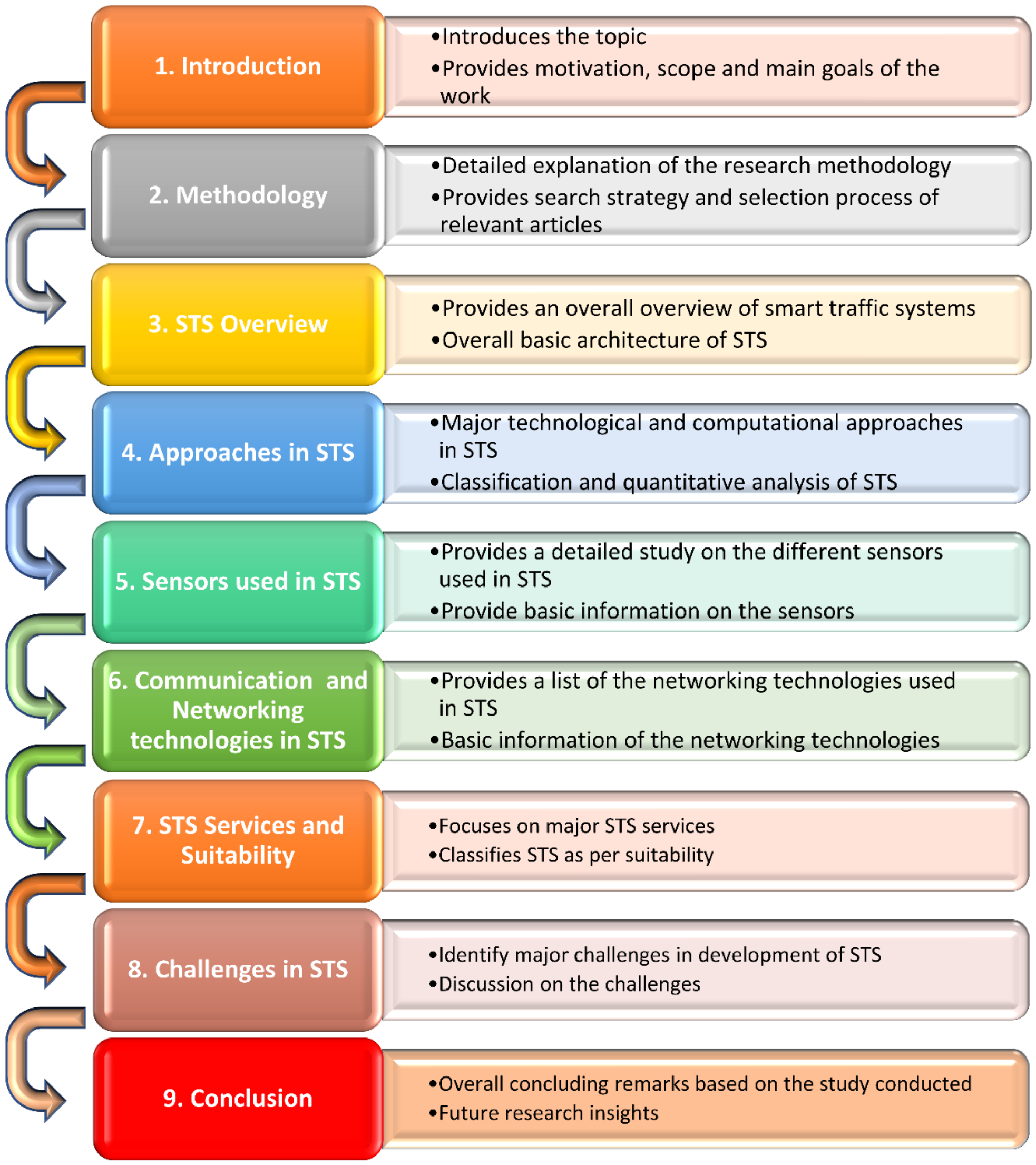

Figure 1. Outline of the paper.

## 2. Methodology

This section describes the methodology followed in this research to find out about previous works, the selection of relevant research materials, and their review process. This research gathered published articles from reputed publishers (IEEE Xplore, Springer Link, ScienceDirect, MDPI, ACM Digital Library, and Hindawi, etc.). The method of searching, finding, selecting, and reviewing published articles is divided into three major phases: preparation, evaluation, and outcome (Figure 2). Descriptions of these three major methodology phases are provided in the following sub-sections.

### *2.1 Preparation Phase*

The preparation phase is the primary stage of the research that is initiated by identifying research objectives, keyword generation, and formulating research questions. To formulate the main objective of this research, the following research questions were established:

- What are the strategies and tactics to establish a smart traffic system?
- What are the main approaches to developing a smart traffic system (STS)?
- Which network tools will be implemented to develop this system?
- What type of sensors will be used?
- What are traffic detection technologies with their advantages and disadvantages?
- What is the qualitative analysis of smart traffic systems with their goals and scenarios?

These are the main themes of the paper. The whole review will be based on these questions. After generating the research keywords and questions, the primary search string is formulated.

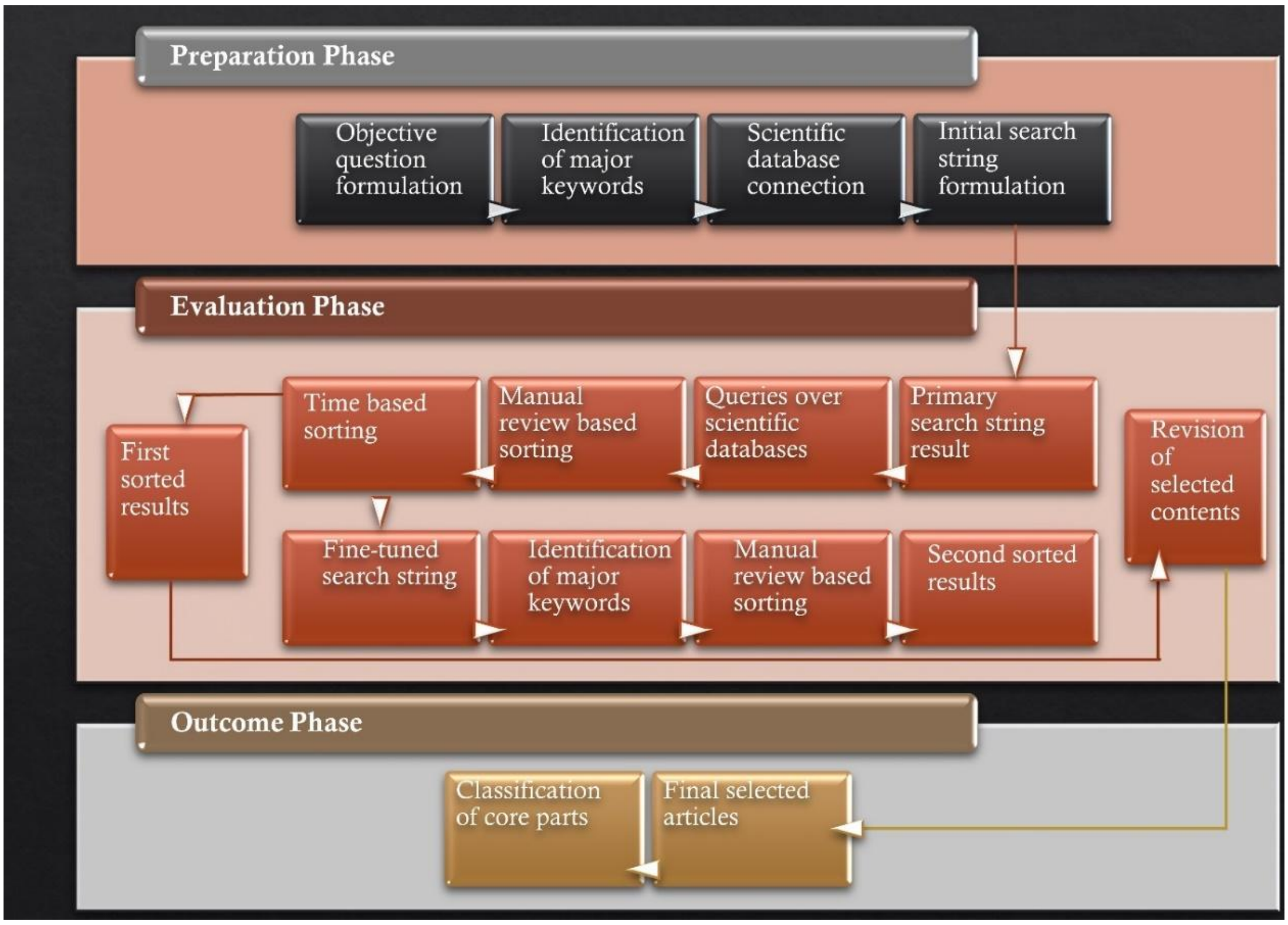

Figure 2. Review methodology.

*2.2 Evaluation Phase*

The search string formulated in the preparation phase is used to find STS-related research works. Following the initial search through the designated scientific databases, the materials are reviewed and sorted to eliminate any potentially irrelevant research works. Then, a second search is conducted based on a fine-tuned search string. Finally, a final revision of the selected contents has been conducted to select the most suitable research materials for this review work. During the evaluation step, several research articles were discovered. However, only research items from the previous 11 years were picked. The selected articles from various online sources were manually sorted by evaluating their quality. A paper is chosen for this step of sorting if it contains the keywords and could give details that could address the set of questions about the study's goal. A paper failing to get connected to the research goals was not selected for review.

*2.3 Outcome Phase*

All the research papers associated with the research have been combined with individual results and searches in this last step of the paper search by various documentation and data tables. The data was then utilized to construct the main body of this report and identify technological trends in smart traffic systems. After that, the selected resources will be applied for final documentation.

In the evaluation stage, there are many papers published by many different publishers. After rigorous selection and sorting, 61 papers are selected. Based on the review, two graphical representations are illustrated in Figures 3 and 4. Figure 3 shows the year-by-year basis of research publication. The paper was taken between the years 2010 and 2020. Here, recently published papers are the main priority. Figure 4 provides the publisher-wise research paper distribution. It is proved from Figure 3 and 4 that recent good quality papers from reputed publishers are considered.

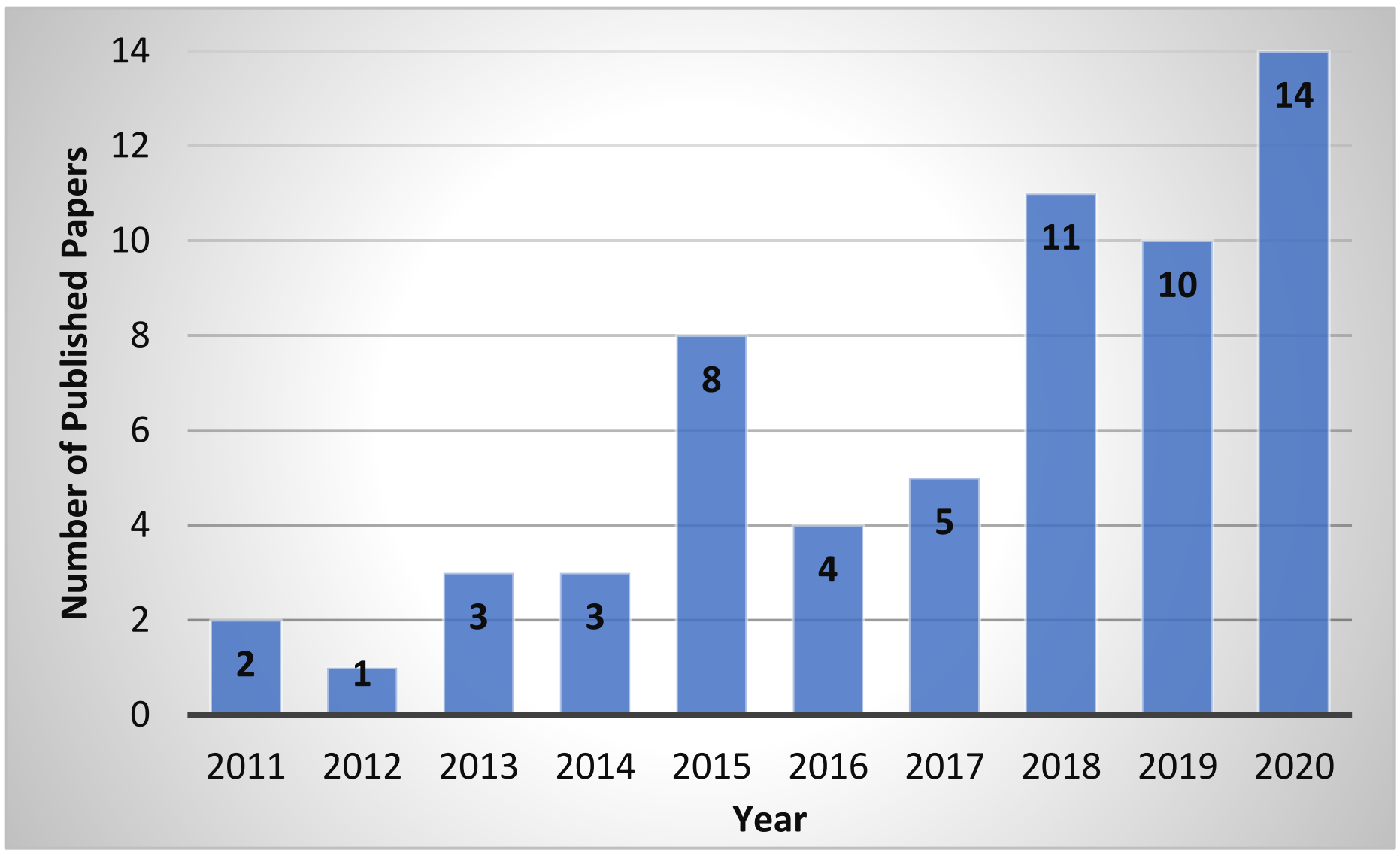

Figure 3. Year-wise distribution of selected research papers for review.

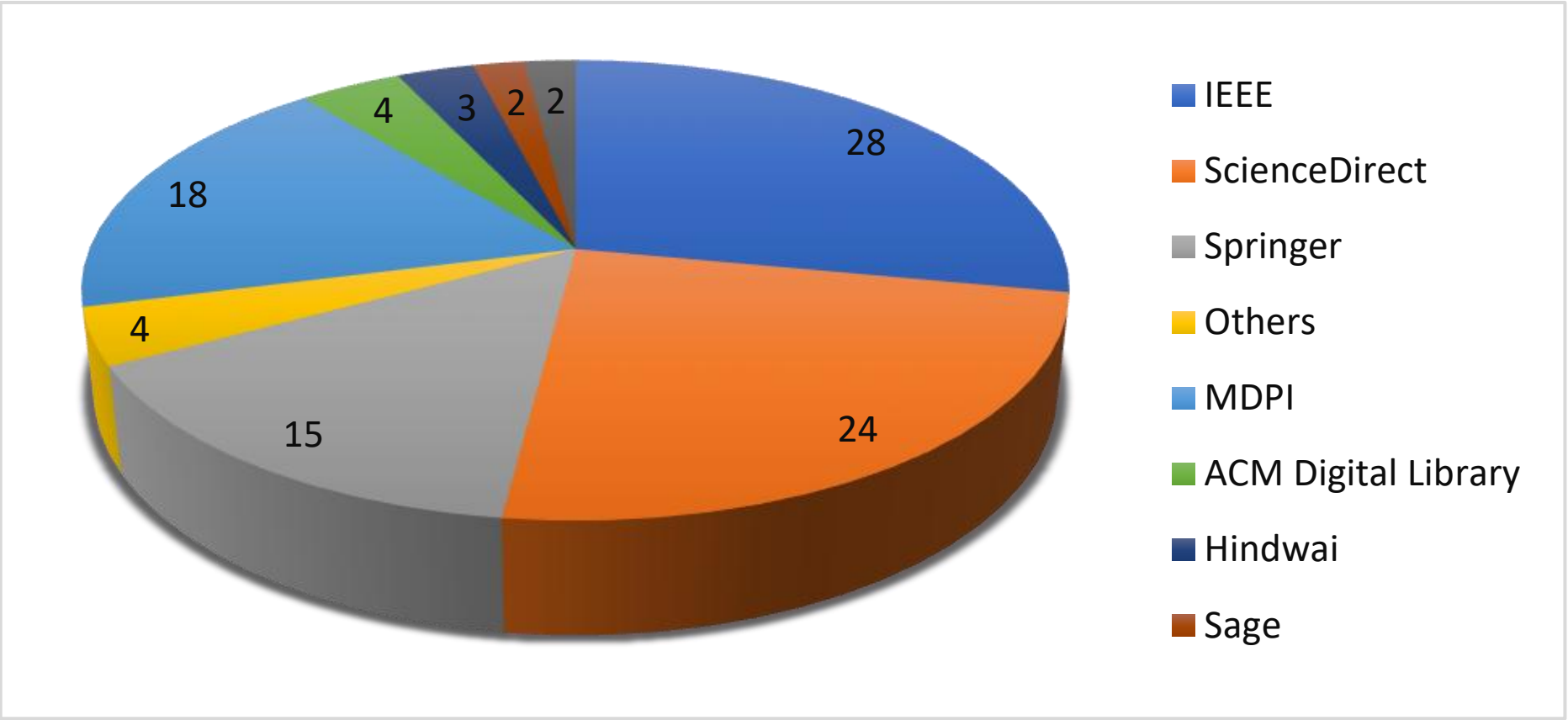

Figure 4. Publisher-wise distribution of selected research papers for review.

## 3. Overview of STS

Figure 5 provides an overview of modern STS. STS can be divided into three major parts:

- Data Collection
- Data Storage and Processing
- Service Delivery

Data collection: All the traffic-related data are collected from various built-in sensors, such as speedometers, odometers and global positioning system (GPS), roadside sensors, in-road sensors, roadside units (RSU), traffic lights, publicly available web sources, and participatory network vehicles. Then, the collected data will be forwarded to the traffic management center, or it should be shared among the neighbors. At the same time, participatory sensing networks and web resources that are publicly available can be used by STS to upgrade the validity of services that are provided, since traffic data is collected from these sources.

Data Processing: Data collected in the Data Collection part is used in this stage to get stored, analyzed, and processed. Different data analysis methods are being used for the effective utilization of data. Moreover, data is sorted and stored as per different types so that it becomes easier for the traffic management center to utilize them to provide the required services.

Service Delivery: This stage uses the processed data to provide specific services based on the system's capacity and needs. Moreover, the goal of this part is to enhance overall STS efficiency. Traffic congestion identification, traffic signal management, accident warnings, emergency condition management, traffic update, road guidance, and many other services are available.

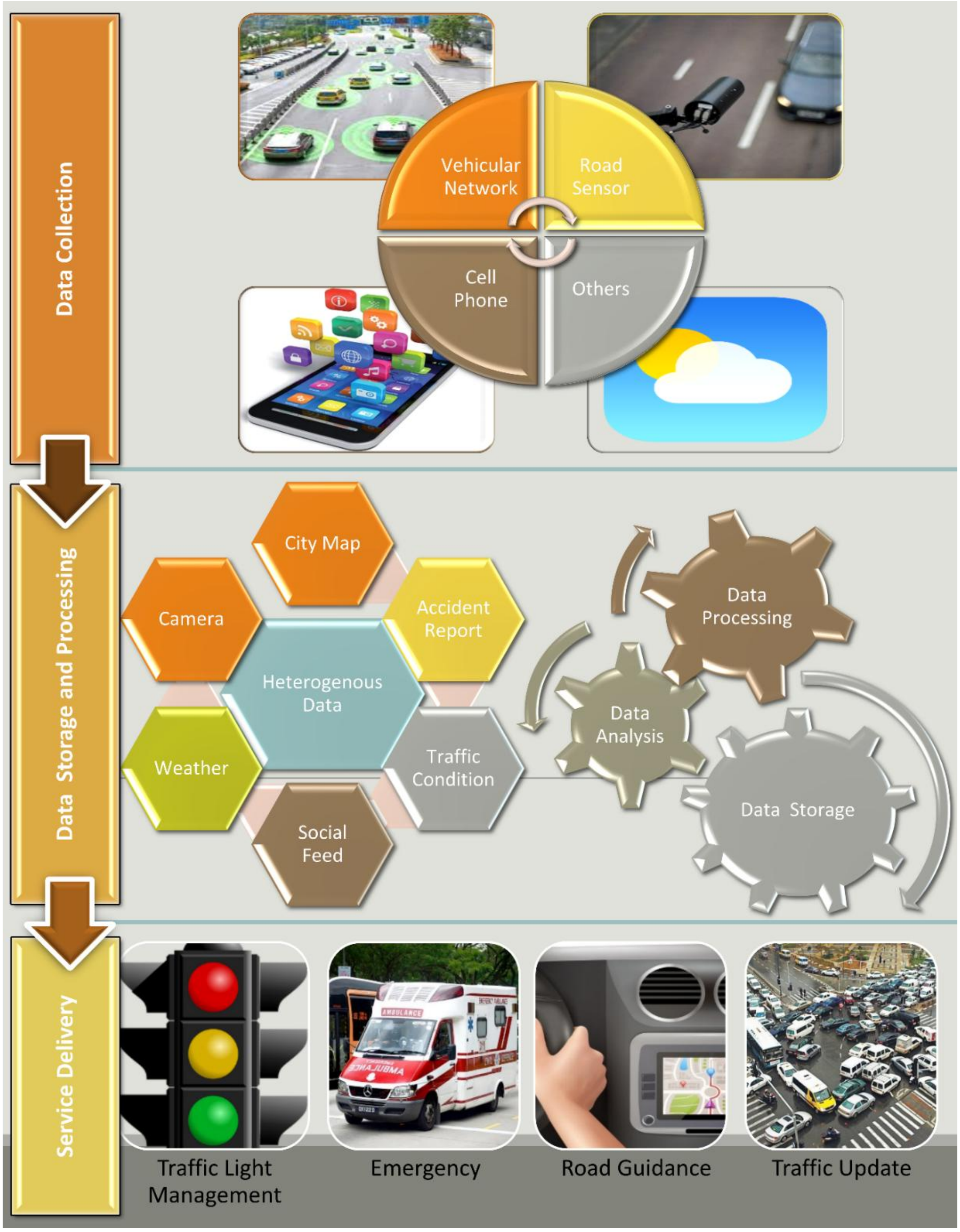

Figure 5. STS Overview.

## 4. Approaches in STS

Details on different STS approaches are provided in sections 4.1-4.11. Section 4.12 provides a summary of STS approaches in various existing research works.

### *4.1 Cellular-based STS (CDMA, GSM, and GPRS Technologies)*

A cellular network is a wireless communication network where multiple user nodes are connected to each other without any physical medium. In this type of network, large areas are divided into small sectors known as cells and each cell contains at least one transceiver module which provides all the nodes available in that cell's network coverage. Cellular networks provide an easy and cheap way for transmitting audio and visual data over long distances. By tapping into this existing network system, it is possible to set up an efficient communication system among the vehicles on the road. Vehicles will be able to communicate on the road and provide information about road congestion, traffic situations, or any emergencies such as accidents or medical needs. Additionally, cellular-phone managers, which are required to record data on the use of these devices for legal or charging purposes, end up with increasingly informative databases. As such, every time a portable station is utilized for calling or sending an SMS, the administrator can record the call highlights which include a timestamp, the base station's identifier with which the client is associated, and information (call span, volume of information traded) [27]. In cell-based methodology-based STS, no exceptionally rare gadget is needed for vehicles, and no foundation is required to be installed around the streets. Therefore, it is less expensive for practical implementation. It is quicker to set up, simpler to introduce and needs less support [89]. Long-range cellular communication can be used to provide early warnings and out-of-sight communication.

*4.2 GPS-based STS*

The Global Positioning System, also known as GPS, is a satellite-based ground navigation system that provides real-time data about the location and speed of an object. Over the years, the GPS system has gained tremendous popularity for navigation purposes. Almost all modern vehicles and mobile phones feature a GPS module that can be used for tracking purposes. By collecting data from GPS modules of the vehicles or from the mobile phones used by the passenger in a certain area, the traffic situation could be identified. GPS is also used for in-vehicle navigation and driving assistance purposes where it can provide data on possible traffic jams or accidents by analyzing the information from other GPS devices in that area. Even though GPS data can provide relevant information with a high degree of quality, investigators decided to analyze the prospect of extracting trip data from GPS data in the mid-1990s. There are still issues with these GPS devices. Reaction shortage to device signals or getting access to the GPS signal in regions such as buildings, caves, tunnels, and city tunnels are some of them. They also use a web-based diary system or a Digital Elevation Model (DEM) to get more details on transport services and trip purposes [90]. Inside buildings, automobiles, and city tunnels, this device can pick up GPS signals (where other infrastructure blocks GPS signals). Technology on this site has expanded, and the receiver is now extremely sensitive. As a result, smartphones will be able to get more precise GPS data [91]. If the advances of naturally determining individual outing information can likewise be accomplished with higher exact outcomes, GPS information assortment through cell phones may turn into the principal technique for individual outing information assortment with cheap expenses and the least weight on respondents in the future [92]. The only drawback is that physical obstructions such as buildings, trees, bad weather, or sometimes even electrical signals can interfere with the efficiency of the GPS.

*4.3 Wireless Sensor Network (WSN) based STS*

WSN means a group of sensors that are wirelessly connected. WSN consists of both data acquisition and distribution networks. The network will be controlled and regulated from a station [93]. It is run by its own power preserving and routing protocols that can be configured in many topologies [30,94]. WSN-based sensor nodes added a variety of sensors that are linked to monitoring different ingredients of the environment. WSN has been shown to be the ideal solution for traffic-related problems because it is resilient, super easy to use, and a well-designed application domain that needs low power consumption, cheap cost, and maintenance.

*4.4 Internet of Things (IoT) based STS*

The internet of things (IoT) has emerged with the notion of smart cities. The physical infrastructures of the cities are equipped with smart gadgets, which are continuously creating multidimensional data in various locations. The idea behind IoT is to create a communication channel among all the electrical and electronic devices used in houses, offices, vehicles, or elsewhere to acquire data and analyze it to provide better user experience and create a smart system. The rising IoT market is gaining traction as operators, vendors, manufacturers, and businesses realize the opportunities it presents [95-96]. Traffic jams are a serious fact that is increasing in tandem with city development. Smart traffic infrastructure is a very essential component of any project for a smart city. Intelligent transportation systems along with integrated components like adaptive traffic signal control, highway management, urgent management services, and roadside units make up a smart traffic system. All ingredients of traffic, including tunnels, roads, bridges, vehicles, traffic signals, and even drivers are included in IoT-based STS [35,36]. Such systems gather data on real-time traffic and take the required steps to ignore or decrease any social topic that may arise because of road congestion. By accessing real-time traffic maps, users will benefit by selecting the best possible route in order to save effort and time. Based on sensor data from monitoring equipment deployed on motorways and urban roads, popular mobile applications such as Google Maps and Apple Maps reliably forecast traffic congestion for urban roadways. These application providers make deals and agreements with various transportation agencies to collect traffic data. Most of the traffic monitoring equipment is placed on city roadways by transportation and road authorities. These application providers (e.g. Google application programming interface) offer updates on traffic jams. Also, these apps use location-based services and crowdsourcing to enhance traffic-density forecasts.

*4.5 Multi-Agent System (MAS) based STS*

A multi-agent system is a computerized process that employs multiple intelligent mediators to solve problems that are otherwise impossible to solve by using a single intelligent agent. In traffic systems, multiple agent-driven systems could help to reduce congestion, avoid accidents, and decrease travel time by figuring out the fastest route. Also, it contributes to decreasing carbon emission by combining real-time traffic data and past travel preferences of the passengers. However, the development of MAS-based STS is facing a variety of complicated challenges which must be addressed, including agent coordination, learning, and security [97]. Because of rapid urbanization, growing motorization, and population expansion, traffic jams have been increasing across the world. Without a doubt, MASs have marked a true milestone in the world of research in the previous decade by bringing together both academics and practitioners from many and sometimes disparate fields [98]. One of the most difficult tasks for traffic light management systems is to be able to process all the information that is generated in their environment and make proper decisions using it. MAS can provide important benefits in solving these problems, by offering the probability of dividing the problem into smaller portions so that the scalability and efficiency of the system will be improved.

*4.6 Machine learning (ML) based STS*

Machine learning is an area of artificial intelligence that works manually using trained models and input provided during the training period. Machine learning uses the data gathered from past interactions and focuses on analyzing and identifying patterns and arrangements in the dataset to improve the existing system without any external human influence. It creates a mathematical model of the known sample data for making estimates or judgments [99]. To separate the traffic area express, ML-based STS dissects the information from the Traffic System [100]. ML can be utilized for traffic congestion forecasting. Moreover, STS dependent on ML can estimate STS inhabitance for the following days, weeks, or even months and give a unique evaluation plan. Frameworks dependent on ML can screen explicit courses and provide an effective solution for the STS framework.

*4.7 Deep Learning (DL) based STS*

Deep Learning (DL) is a subset of ML and Artificial Intelligence that copies the human cerebrum's information handling and highlights extraction to settle on decisions [101]. Computer vision, object perception, visual searches for retail sectors, self-driving smart vehicles, speech recognition, home security, robots, and natural language processing are all utilizing DL systems. Rather than utilizing customary sensors, DL calculations perceive blankly involved and uncommon traffic areas in an STS, decreasing the number of sensors and cameras needed by the framework. The collection of traffic data is an important component of understanding and improving safety [102]. However, DL process requires large quantities of data in order to provide efficient results. Because of the vast quantity of data involved, the cost of training a system increases substantially. Additionally, Deep Learning requires expensive computing systems and is quite complex to set up.

*4.8 Neural Network (NN) based STS*

NN is an assortment of calculations that utilizes a strategy that emulates human cerebrum movement to remove qualities and fundamental connections from information sets. In STS, NN is used for constant video information tag acknowledgment. The neural network is also utilized for successful license plate number recognition [103]. To decide if a traffic area is involved, CNN and machine vision are utilized. CNN may likewise give data on traffic conditions along a different course. Because of its capacity to handle multidimensional data, flexible model structure, great generalization, learning ability, and flexibility, NN is a quite preferred method for traffic prediction [104]. The only disadvantage of the Neural Network system is its 'Black-Box' nature which simply means that human analysts cannot check or alter why the algorithm came up with the result regardless of whether it is right or wrong. Thus, it creates the dilemma of whether to trust the output given by a NN in a sensitive sector such as a traffic system.

### *4.9 Fuzzy Logic-based STS*

Fuzzy logic is a style of thinking similar to human reasoning. Fuzzy logic control allows for the manipulation of linguistic and inexact traffic data as a valuable tool in the construction of traffic light timing schemes. With the fuzzy input variables of arrival and queue, traditional fuzzy traffic controllers modify the extended time of the green phase. In STS, fuzzy logic is utilized to forecast traffic system occupancy [105]. However, without testing the prediction result with real-time data, the accuracy of the Fuzzy logic prediction model will be low. To adjust the length of traffic signals dynamically, the image subtraction approach, which requires a shorter time to detect cars, is combined with fuzzy logic [106]. So, fuzzy logic with sensors or machine vision improves the accuracy of the whole system.

### *4.10 Bluetooth-based STS*

Bluetooth is a short-range wireless-communication technology that allows data transfer. Automated traffic control is generally included in a smart traffic system which is entirely dependent on Bluetooth technology. The Bluetooth sensor systems for STS rely on many Bluetooth-enabled devices being used [48, 107, 108]. Regular STS, which does not establish an automated traffic control system, will require extra sensors and methods to obtain various smart traffic amenities. Easy availability and low cost make Bluetooth an ideal system for short-distance communication, such as between sensors and processing units. A Bluetooth system can also be employed to establish communication between vehicles. But, due to the short range, it is not very feasible most of the time. The crowd-sensing approach is used by many smart traffic systems to acquire information about traffic places in a given location. Smartphone sensors, accelerometers, gyroscopes, magnetometers, and GPS and apps are used to collect traffic data.

### *4.11 Vehicular Ad-Hoc network (VANET) based STS*

The concepts of mobile ad hoc networks (MANETs) – the natural development of a wireless network of mobile devices – are applied to the realm of cars to produce vehicular ad hoc networks (VANETs). In the realm of wireless communication, VANET has risen as one of the most significant study areas. VANETs have a high expectation of improving road safety by providing citizens with a variety of amenities [109]. VANET uses a short-range communication system that is dedicated, and it allows communication between moving vehicles (DSRC) wirelessly [110]. VANET is an important element of the architecture of an intelligent transportation system [111].

### *4.12 STSs Classification Based on Technological Approaches*

A summary of STS classification based on technological approaches is presented in Table 1. IoT-based STS is the widely used approach. WSN, GPS, Fuzzy Logic, and MAS are also frequently used approaches in STSs. Cellular and Bluetooth-based approaches are the least utilized ones.

Table 1. STS classification based on technological approaches.

| References | WSN | IoT | Multi-Agent | GPS | Cellular | Bluetooth | Machine Learning | Deep learning | Neural Network | Fuzzy Logic | VANET |
|---|---|---|---|---|---|---|---|---|---|---|---|
| [28] | + | + | | | | | | | | | |
| [29] | | + | | | | | | | | | |
| [30] | + | | | + | | | | | | | |
| [31] | + | + | | | | + | | | | + | |
| [32] | | + | | | | | | | | + | |
| [33] | + | + | | | | + | | | | | |
| [34] | | | | | | | | | | + | |
| [35] | | | | | | | | | | | |
| [36] | + | + | | | | | | | | | |
| [37] | | | | + | | | | | + | + | |
| [38] | | + | | + | | + | | + | + | | |
| [39] | | | | | | | | | + | + | |
| [40] | | + | | | | | | + | + | | |
| [41] | | + | | + | | | | | | | |
| [42] | + | + | | | | | | | | | |
| [43] | + | + | + | | | | | + | + | | |
| [44] | | + | | + | | + | | + | + | | |
| [45] | | + | | | | | | + | + | | |
| [46] | | | | | | | | | + | | |
| [47] | | | + | | + | | | | + | + | |
| [48] | | + | | | | + | | + | | | |
| [49] | | | | + | | | | | | | + |
| [50] | | + | | | | | | | | | |
| [51] | | | | + | + | + | | | | | |
| [52] | | | + | + | | | + | | | | |
| [53] | | + | + | | | | | | | | |
| [54] | | | | + | | | | | | | |
| [55] | | | | | | | + | | | | |
| [56] | + | + | | | + | | | | | | |
| [57] | | + | | | | | | | | | |
| [58] | | | + | | | | | | | | |
| [59] | | | + | | | | | | | | |
| [60] | | + | | | | | | + | | | |
| [61] | | | | | | | | | | + | |
| [62] | | + | | | | | | | | | |
| [63] | + | | | | | | | | | | |
| [64] | | + | | | | | | | | | |
| [65] | + | + | | | | | | | | | |
| [66] | | | | | | | + | | | | |
| [67] | | | + | | | | | | | | |
| [68] | | | | | | | | | | | + |
| [69] | | | | + | | | | | | | |
| [70] | | | | | | | | | | | + |
| [71] | | | | + | | | | | | | + |
| [72] | | | | | | | | | | | |
| [73] | | | | | | | | | | | |
| [74] | | | | + | | | | | | | |
| [75] | | | | | | | | | | | |
| [76] | | | | | | | | | | | |
| [77] | | | | | | | | | | | + |
| [78] | | | + | | | | | | | | |
| [79] | | | | | | | | | | + | |
| [80] | | | | | | | | | | + | + |
| [81] | | | | | | | | | | | |
| [82] | | | + | | | | | | | + | |
| [83] | | | | | | | | | | | |
| [84] | | | | | | | | | | | + |
| [85] | | | + | | | | | | | | |
| [86] | | | | + | | | | | | | |
| [87] | | | + | | | | | | | + | |
| [88] | | | + | | | | | | | + | |
| Total | 10 | 22 | 12 | 13 | 3 | 6 | 3 | 7 | 9 | 12 | 7 |

## 5. Sensors Used in STS

Details on different sensors used in STS are provided in sections 4.1-4.10. Section 4.12 provides a summary of the sensor topologies and their specifications.

### *5.1 Camera*

Cameras are the crucial sensing device on the street to collect real-time statistics. The application of cameras has become more popular with the increasing use of auto-driving systems in modern vehicles. In STS, these cameras are used for screening roadways, congestion, and vehicle identification. Every camera is controlled on a clockwise foundation via the workforce at the city's Traffic Operations Centre (TOC) [38-39]. However, the cost of implementation and the storage required limits its widespread utilization.

### *5.2 Cellular Sensor*

Sensors installed inside smartphones are called cellular sensors [112-113]. Cellular sensors of multiple smartphones together create a Cellular Network. Among the various sensors present in smartphones, magnetometers, gyroscope and accelerometer are the main sensor technologies that are useful for STS. Motion, direction, and orientation of a user can be extracted from these sensors which can be used to detect traffic congestion in a given area.

### *5.3 Infrared (IR) sensor*

Theoretically, almost all objects or animals emit some amount of IR radiation. IR sensor is a digital sensor that detects and measures the infrared radiation in its nearby ambiance. It can also detect the presence of people when they are in front of it [114]. The main purpose of using an IR sensor is to detect motion in its surroundings. Usually, in STS, the sensors are installed underneath the pavement of the street to detect vehicular movement.

### *5.4 Inductive Loop Detector*

An inductive Loop Detector functions similarly to a metal detector [115]. It produces an electromagnetic field and when a vehicle enters this field, the detectors can detect its presence and classify the vehicle. In STS, Inductive Loop Detectors are found at intersections, parking lots, driveways, or car gate systems [116]. With some computational processes, data obtained using this sort of sensor can provide various STS services.

### *5.5 Agent*

An Agent consists of processors, sensors, and some other tools within it. An agent is a part of an extensive system (such as a multi-agent system) that is responsible for generating, processing, and transmitting data. Multi-agent structures might be a useful resource to simplify the harmony of STS [47-48]. An Agent is a term that is used to indicate a sensible actor that interacts with its neighborhood via methods of actuators and sensors. A multi-agent system is a machine produced by different retailers that is capable of obtaining a goal that is hard or not possible to obtain through a monolithic gadget.

### *5.6 Ultrasonic sensor*

Ultrasonic sensors are generally used as proximity sensors. The sensor emits and receives ultrasonic signals by which it detects and measures the distance between objects. These may be seen in cars with anti-collision systems and self-parking technology. It can measure distance easily. It is perfect for measuring the distance between dynamic or static objects. It can also measure the distance in the air through a non-contact technique. It is easy to use and reliable and can measure distance without any damage [117]. Ultrasonic sensors are also utilized in robotic impediment detection systems, as well as in the manufacturing era. These sensors discover the presence of an automobile using the reflection of ultrasonic waves transmitted from ultrasonic transmitters/receivers mounted above the street [118].

### 5.7 Microwave Radar

Microwave Radar detects an object touring near or far from it by transmitting a signal and receiving the signal echoed from the target object. The echoed signal is then further processed to find the target object's velocity [119]. This sort of sensor performs well in multiple-lane scenarios and works fine in bad weather conditions. But some microwave radars (i.e. continuous wave doppler sensors) are incapable of detecting immobile vehicles. So, this kind of sensor is mainly used for counting vehicles, accessing their speed, and classification of vehicles in a particular traffic environment.

### 5.8 Magnetometer

A Magnetometer can detect an object by sensing the change in electromagnetic fields in the surrounding environment. The magnetometer can sense the horizontal and vertical components of Earth's magnetic field [120,121]. A Magnetometer can be used at the side of a 3-axis accelerometer to produce orientation impartial accurate compass heading [121]. A Magnetometer can handle more traffic volume or pressure than regular loop detectors and foul weather conditions do not hamper its operation. But the installation and maintenance of magnetometers for traffic management are costly as it requires boring pavements and roads.

### 5.9 Vehicular Sensor Network

A vehicular sensor network consists of a bunch of wired or wireless or networked infrastructure. It connects to the internet or any computer network through which it transfers data for analysis and use. Sensor network nodes can sense cooperatively and regulate the environment. For proactive stacking of traffic information in city environments, vehicular sensor networks are emerging day by day. Such a sensor network has the ability to process sensed data (e.g., license plate recognition), route messages to other vehicles, and sense incidents (e.g., imaging from streets) [122-123].

### 5.10 Radio Frequency Identification (RFID) sensor

In RFID, digitally encoded data is placed in RFID tags or smart labels are intercepted by a receiver via radio waves. RFID and barcode scanning is almost the same as both collecting data from a tag or label and store it in a database [124]. The reading range is highly dependent on the tag antenna. RFID sensors can be useful to detect and identify vehicles in STS [125].

### 5.11 Summary of Sensors Utilized in STSs and Their Specifications

An outline of different sensor topologies used in STSs is given in Table 2. It is observed from Table 2 that the Camera is the most preferred sensor topology in SPS [126]. It is due to the fact that the Camera can provide multiple types of traffic-related information. Agent and Vehicular Sensors are also widely used sensor topologies. Table 3 highlights the summary of sensing technologies used in STS.

Table 2. Different sensor topologies are used in different STSs.

| Reference | Camera | Cellular Sensor | Infrared IR Sensor | Inductive Loop Detector | Agent | Ultrasonic Sensor | Microwave (RADAR) | Magnetometer | Vehicular Sensor | RFID sensor |
|---|---|---|---|---|---|---|---|---|---|---|
| [28] | + | | | | | | | | | |
| [29] | + | | | | | | | | + | |
| [30] | + | | | | | | | | | |
| [31] | + | | | | | | | | | |
| [32] | | | + | | | | | | | |
| [33] | + | | | | + | + | | | | |
| [34] | | | | | | | + | | | |
| [35] | + | | | | | | | | | |
| [36] | + | | | | | | | | | |
| [37] | | | | | + | | | | | |
| [38] | + | | | + | | | | | | |
| [39] | + | | | | | | | | | |
| [40] | + | | | | + | | | | | |
| [41] | + | | | | | | | | | |
| [42] | + | + | | | | | | | | |
| [43] | | + | | | + | | | | | |
| [44] | + | | | + | | | | | | |
| [45] | + | | | | | | | | | |
| [46] | + | | | | | | | + | | |
| [47] | | | | | + | | | | | |
| [48] | + | | | | + | | | | | |
| [49] | | + | | | | | | | | |
| [50] | | | | | | | | | | + |
| [51] | | | | | | | | | | |
| [52] | + | + | | | | | | | | |
| [53] | | + | | | | | | | | |
| [54] | | | | | | | | | | |
| [55] | | | | | | + | | | | |
| [56] | | | | | | | | | + | + |
| [57] | | | | | + | | | | | |
| [58] | | | + | | | | | | | |
| [59] | | | | | + | | | | | |
| [60] | + | | | | | | | | | |
| [61] | | | | | | | | | | |
| [62] | + | | | | | | | | | |
| [63] | | | | | | | + | | | |
| [64] | + | | + | | | | | | | |
| [65] | | | + | | | | | | | |
| [66] | | | | | | | | | | |
| [67] | | | | | | | | | + | |
| [68] | | | | | | | | | | |
| [69] | | | | | | | | | + | |
| [70] | | | | | | | | | + | |
| [71] | + | | | | | | | | + | |
| [72] | | | | | | | | | + | |
| [73] | | | | | | | | | + | |
| [74] | + | | | | | | | | | |
| [75] | | | | | | | | | + | |
| [76] | + | | | | | | | | + | |
| [77] | | | | | | | | | | |
| [78] | | | | | + | | | | | |
| [79] | + | | | | + | | | | | |
| [80] | | | | | | | | | + | |
| [81] | | | | | | | | | + | |
| [82] | | | | | + | | | | | |
| [83] | | | | | | | | | + | |
| [84] | | | | | | | | | + | |
| [85] | | | | | + | | | | | |
| [86] | + | | | + | | | | | | |
| [87] | | | | | + | | | | | |
| [88] | + | | | | + | | | | | |

| Total | 25 | 5 | 4 | 3 | 14 | 2 | 2 | 1 | 14 | 2 |
|---|---|---|---|---|---|---|---|---|---|---|

Table 3. Outline of traffic sensing technologies.

| Technology | Particular Equipment | Advantages | Disadvantages | Principles |
|---|---|---|---|---|
| Aerial/Satellite Imaging | Helicopters, Drones | • Easy to detect traffic congestion and accidents.<br>• Can cover a large area with<br>• Can provide a bird's eye view of traffic. | • It often becomes difficult to analyze the traffic condition from aerial images.<br>• Obtaining traffic updates is expensive and time-consuming.<br>• It has massive databases and image recognition takes a long time. | Utilization of unmanned or manned vehicles that can capture the aerial view of streets, roads, and highways. Captured images and videos are analyzed to provide STS services |
| Inductive loop | Sensors installed underneath the roads, several electronic units and a control panel | • Has the flexibility to be applied in a wide range of applications.<br>• Inert to a bad environment condition.<br>• Can provide accurate data for analysis | • Need to cut existing roads to insert the sensors underneath the roads.<br>• Numerous circles are needed to cover an area.<br>• The recognition precision drops with vehicle classes. | The sensor senses the loop inductance metal objects of vehicles |
| Infrared | Camera with multi spectrum operation abilities | • Information such as position, vehicle class and speed can be accurately determined | • Performance degrades in poor environment conditions.<br>• Requires installation of several devices.<br>• The installed devices need regular maintenance and cleaning | The transmitter part sends low power infrared waves. The reflected waves from vehicles are captured and analyzed to understand the presence of vehicle |
| Magnetometer | Micro loop probes, Magnetic probe detector and a control unit | • Can handle high traffic congestion<br>• Less sensitive to environmental changes | • Difficulty in installment under roads.<br>• Difficulty of replacing damaged sensors since sensors are installed underneath the roads. | Vertical and horizontal components of the Earth's magnetic field is measured by means of sensors installed in Magnetometers |
| Microwave RADAR | Electromagnetic transmission and receiver antenna, processor | • less sensitive to environmental changes.<br>• Has the ability to operate in multiple lanes. | • Consistent wave Doppler sensors do not have the ability to detect motionless vehicles standing in traffic jams. | The electromagnetic transmitter antenna emits microwaves. The reflected waves from vehicles are detected and process for extracting traffic information |
| RFID | RF antenna, RFID tags, RFID tag reader, computer as processor | • RFID tags are cost efficient and easy to install in vehicle<br>• RFID can provide exact identification of vehicles | • Only vehicles equipped with RFID tags are detected. So, it is unable to provide a total scenario of traffic condition | Utilizes Radio waves to detect and identify vehicles. The vehicles carry RFID tags and RFID readers are installed in selected locations |
| Ultrasonic | Oscillator, amplifier, Ultrasonic receiver and transmitter | • Has the ability to monitor several lanes simultaneously.<br>• Fast response with low processing time for congested areas | • Performance is influenced by natural conditions.<br>• Requires | The sensor emits ultrasonic waves which gets reflected from the vehicles. The reflected waves are captured by which a vehicle can be tracked. Location between object and sensor is understood by the tracking the time lapse of the ultrasonic wave |
| Camera | Color camera along with an image processor | • Can monitor a wide area<br>• Provide several types of useful data | • High installation cost and data rate.<br>• Performance dependent on environmental factors such as rain, shadow, darkness, etc. | Camera based sensing method involves a network of camera and image processing units. Processed data are sent to the control center by wired and wireless mediums. |

**6. Communication and Networking Technologies Used in STS**

One of the most vital parts of an STS is its communication and networking tools. The communication and networking tools of an STS work as a medium between the sensor nodes and the processing unit. Nowadays, communication between the sensor nodes and the processing unit is done via various wireless communication and networking technologies. Wireless communication is accomplished by a simple sensor node but the connection range for communication depends on its technology, which can range from meters to kilometers. For example, Bluetooth can cover less than 10 meters, Zigbee less than 75 meters, and Wi-Fi less than 100 meters. Also, WiMAX can have a range of fewer than 10 kilometers. Wi-Fi, Bluetooth, Zigbee, GPRS, GSM, RFID, and other wireless networks are examples of communication protocols. A tabulated discussion is provided in Table 4 discussing various wireless communication technologies used in various STSs.

Table-4. Specifications of several communication and networking protocols used in STS

| | **Bluetooth** | **GPRS** | **GSM** | **RFID** | **Wi-MAX** | **Wi-Fi** | **ZigBee** |
|---|---|---|---|---|---|---|---|
| **Frequency** | 2.402 to 2.480 GHz | 850/900/ 1800/1900 MHz | 933 - 960 MHz | low-125 kHz, high- 13.56 MHz, ultra- high- 860-960 MHz | 5091 MHz to 5150 MHz | 2.4 GHz to 5 GHz | 2.4 GHz |
| **Standard** | IEEE 802.15.1 - 2002 | - | - | - | IEEE 802.16e - 2001 | IEEE 802.11 - 1997 | IEEE 802.15.4 - 2006 |
| **Description** | Mechanical determination for WPAN empowers voice and information transmission between various gadgets through a secure, universally free radio connection (2.4 GHz). | Amplified GSM for packet information correspond- ence | Commonplace framework for communication through mobile phones including digital technology | Utilizes radio waves to recognize objects conveying labels. | Standard for information transmission by means of radio waves. | Framework of remote data broadcast over computational networks. | Determination of a bunch of complex remote correspondence conventions for use with low utilization computerized radios, in view of WPAN standard IEEE 802.15.4 |
| **Feature** | Law power version accessible | High asset usage, short get To time | Huge scope, High capacity and transmission quality. | Low cost | Rapid and serve a number of clients. | Fast what's more, universality | Cross section organizations, Different convention accessibility. |
| **Throughput** | 1 Mbps | 56–144 Kbps | 9.6 Kbps | 9.6–115 Kbps | <75 Mbps | 11/54/300 Mbps | 250 Kbps |
| **Range** | 30 feet (10 meters) | dependent on cellular arranged service provider | dependent on cellular arranged service provider | Up to 3 m | <10 km | <100 m | <75 m |

**7. STS Services and Utilization Suitability**

Major services provided by different STSs are analyzed and summarized in this section. Moreover, the suitability of STSs is also discussed. The following sub-sections will emphasize these topics. Section 7.1 emphasizes STS services whereas section 7.2 emphasizes STS suitability. Finally, section 7.3 provides a summary table of STSs based on the service provided and suitability.

*7.1 Major STS Services*

Modern traffic systems are not only confined to traffic light management. Rather, the present-day STSs provide a wide range of services. The major STS services are covered in sections 7.1.1-7.1.7.

7.1.1. Accident Detection

Accident detection and early warning of accidents on the road is one of the most important features of a Smart Traffic System. The execution of programmed street accident detection frameworks to give a convenient guide is highly important. Numerous research works have been conducted on this topic. The procedures incorporate accident forecasts utilizing cell phones, vehicular sensors, GPS/GSM-based frameworks, and different AI methods [128,129].

7.1.2. Traffic Congestion Detection

In STS, traffic congestion is detected by utilizing cameras, GPS, and other systems to track the vehicles on the road. For effective traffic light management, it is necessary to detect traffic congestion on roads and highways [130,131]. Therefore, traffic congestion detection has become the major source of concern for traffic control centers. Due to this fact, a wide range of traffic congestion detection schemes have been proposed by various researchers over the course of time. This facility in STS helps users to avoid congestion and offers better traffic management.

7.1.3 Traffic Condition

Slower speed, greater journey times, and increased traffic waiting in line characterize the traffic situation in transportation. Congestion or a highway snarling up occurs when transportation is entirely stopped for a long period of time. The conditions of road traffic must be collected to effectively control travel times. On a single stretch of road, flow, and density can be used to describe the traffic situation. The length of the segment may differ based on the road's geometry. When estimating traffic conditions, various modeling approaches are frequently used [75, 76, 132]. However, because models cannot incorporate all characteristics of the real system, they must be supplemented with observed traffic data such as traffic counts and speed/travel time measurements to provide a good depiction of reality.

7.1.4 Traffic Classification

Traffic classification is the identification and classification of different types of vehicles [133]. Traffic classification is necessary for urban areas and highways to detect vehicles that have emergency issues. For example, the classification and identification of ambulances, fire-fighting trucks, and police vehicles are necessary to provide priority. Moreover, the classification of vehicles in highway toll plazas is necessary to automatically determine tolls based on the type of vehicle.

7.1.5 Data Dissemination

Dissemination is the delivery of data to clients, acquired through a measurable movement. Data dissemination comprises circulating or sending factual information to clients. Traffic dissemination is characterized as a structure to disseminate and accumulate data about the vehicles out and about [134,135]. With such a framework, the vehicle's driver will be given street traffic data that helps driving in circumstances such as foggy climate or tracking down an ideal course in an outing a few miles in length.

7.1.6 Speed Adjustment

Traffic speed adjustment in urban areas and highways is another important service provided by modern STS [136]. In urban areas and highways, vehicles are required to maintain the maximum speed limit for safety purposes [137]. Moreover, speed adjustments are required in school zones, hospital areas, and crowded sites. As a result, several research works have been done on speed adjustment for STS. Modern STSs can help the traffic management authorities by providing information on vehicles that have crossed speed limits and the location of the incident. Additionally, the latest self-driving vehicles are capable of maintaining speed limits automatically, thereby reducing the chance of accidents.

7.1.7 Congestion Avoidance

A congestion avoidance system enables a driver to take a different route to their destination than a route that is congested with traffic. Such schemes try to avoid traffic congestion by suggesting less congested routes to the drivers. In modern cities, the number of vehicles has increased at an alarming rate. However, the existing roads can only be extended to a certain limit because of buildings and other infrastructures. As a result, smart distribution of vehicle routes to avoid congestion in a particular area becomes an effective choice. Due to this reason, several research works have been conducted focusing on congestion avoidance schemes in STS [68,69,138].

*7.2. STS Suitability*

As per suitability of utilization, STS can be classified into two major groups: STS for urban and STS for highways.

7.2.1. STS for Urban

The convergence of financial potential and populace in the metropolitan regions brings about the event of enormous vehicle needs in a spatially restricted region. Therefore, urban traffic conditions incorporate a large number of slow-moving vehicles with a very small distance between them. Moreover, streets in the city areas are packed with numerous citizens. Traffic management in urban areas is concerned with traffic management, speed limitation in populated areas, schools, medical facilities, parking space allotment, congestion alerts, and crowd management. To address these problems, STS in urban areas requires special infrastructure suited to the traffic conditions in highly congested areas. As a result, STS infrastructure and services are a bit different than STS required on highways [69,70,139].

7.2.2. STS for Highway

STS on highways needs to deal with fast-moving vehicles over long distances. Therefore, vehicle identification, speed determination of vehicles, and speed limit adjustment become the major requirements [72,73]. Moreover, STS in highway toll plazas needs to be equipped with vehicle categorization or classification methods and automatic toll collection processes [140].

*7.3. Summary of STS Services and Suitability*

Table 5 provides a summary of different STS services. Moreover, the suitability of different STS research works is highlighted so that it becomes easier to choose the correct STS topology as per system requirements. It can be observed from Table 5 that researchers are more focused on traffic congestion detection so that the traffic lights can be effectively and efficiently managed. Other services such as accident detection, traffic condition awareness, data dissemination, traffic speed adjustments, and congestion avoidance schemes are given a similar sort of priority. Moreover, it is observed that most of the STS research work is suitable for urban areas. It is due to the fact that modern urban areas face more severe traffic management issues compared to highways.

Table 5. STS services and suitability.

| Reference | Services | | | | | | | Suitability | |
|---|---|---|---|---|---|---|---|---|---|
| | Accident Detection | Congestion detection | Aware about traffic condition | Traffic classification | Data dissemination | Speed Adjustment | Congestion avoidance | Urban | Highway |
| [28] | + | + | + | + | | + | | + | |
| [29] | + | | + | + | | + | | + | + |
| [30] | + | + | | + | + | + | | + | |
| [31] | + | | | + | + | + | + | + | + |
| [32] | | | | + | | + | + | + | |
| [33] | + | | | + | + | + | | + | |
| [34] | + | + | + | | + | + | | + | |
| [35] | | + | | + | | + | + | | + |
| [36] | | + | + | + | | | + | | |
| [37] | + | | + | | + | + | + | + | + |
| [38] | + | | | + | + | + | | | |
| [39] | + | | + | + | | | | + | |
| [40] | | + | | + | | | + | + | |
| [41] | + | | | + | | + | | + | + |
| [42] | + | + | | | + | + | | + | |
| [43] | + | + | + | | | + | + | + | |
| [44] | | | | + | + | + | | + | |
| [45] | | + | | + | + | | + | | |
| [46] | | | + | + | | + | | + | |
| [47] | | + | + | | + | + | | + | |
| [48] | + | + | | + | | | | | + |
| [49] | + | + | | | + | + | + | | |
| [50] | | | | + | + | + | | | |
| [51] | | + | | + | | + | + | | + |
| [52] | | | + | | + | | | + | |
| [53] | | | + | | | | | | + |
| [54] | | + | + | | | | + | + | |
| [55] | | + | | | | + | + | + | |
| [56] | + | + | | | | | | + | |
| [57] | | + | | | | | + | | |
| [58] | | + | | | | + | | | + |
| [59] | | + | | | | | + | | |
| [60] | + | | | | + | | | | + |
| [61] | | + | | | + | + | + | | + |
| [62] | + | + | | | | + | + | + | + |
| [63] | | + | + | + | | + | | + | |
| [64] | + | + | | | | | + | + | |
| [65] | + | | | | | + | + | + | |
| [66] | | | + | | | + | | + | |
| [67] | | + | + | | | | | + | |
| [68] | | + | + | + | + | | + | | + |
| [69] | | + | + | | | | + | + | |
| [70] | | + | | | + | | + | + | |
| [71] | + | | | | | | + | + | |
| [72] | | + | | + | | | + | | + |
| [73] | + | + | | | + | | | | + |
| [74] | + | + | | | + | | | + | |
| [75] | | + | + | | | | + | + | |
| [76] | | + | + | | | | + | + | |
| [77] | + | | | | | | + | + | |
| [78] | | | | + | | | | + | |
| [79] | + | + | | | | | | + | |
| [80] | | + | | + | + | | + | + | |
| [81] | + | + | | | + | | | | + |
| [82] | | + | | | | + | | + | |
| [83] | | + | | + | | | | | + |
| [84] | | + | | | + | | + | + | |
| [85] | | + | | | | + | | + | |
| [86] | | + | + | | | | + | + | |
| [87] | | + | + | | | | | + | |
| [88] | | + | | + | | + | | + | |
| Total | 19 | 42 | 21 | 25 | 22 | 29 | 29 | 42 | 17 |

## 8. Challenges, Risks, and Prospective Solutions

STS decreases the journey time of city site visitors by means of ensuring safety with the intention of effective traffic management. Owing to the large amount of traffic in the modern world, STS faces several technological challenges. The major challenges in STS are highlighted in the following sub-sections.

### *8.1. Data Integration*

Due to the enormous number of sensors present in STS, data integration becomes a major challenge. It often becomes difficult to integrate the heterogeneous data obtained from different sources into a single system. To make things worse, several other issues will arise after the heterogeneous data is integrated into the system. Managing and tracking the numerous sensors and devices that are utilized for data generation will pose a significant challenge. The challenges related to data integration are as follows:

- Creating a mechanism for device identification and providing unique identification codes to the devices
- The process of using the identifiers to forward device-specific information
- The method of providing IoT-based identifiers for STS.
- Information generated from human-owned devices and social media posts might have personal information. Therefore, a security mechanism needs to be implemented while integrating this sort of data

### *8.2. Application Domain and Internet Domain Security Issues*

STS wireless networks have two different types: ad-hoc and Wi-Fi based. The ad-hoc part connects the system nodes with each other. IoT-based nodes are simple to penetrate. Therefore, it enables offenders to grab a node to get critical penetration information and resources. On the other hand, the Wi-Fi-based part joins the network to the Internet, which also adds risks to the network for an injection attack. An injection attack affects the decision of the system where the real data might be interchanged with fabricated information.

The ideas to resolve these issues are:

- Implementation of the encryption of the network and access control of the Wi-Fi part.
- It uses different types of different add-ons for Authentication and between nodes verification.

The purpose of the system is to solve transport-related problems. STS consists of many sensors to collect and analyze data continuously. Different devices that are connected with each other can share a large number of confidential data and information via the WWW (World Wide Web) and using a variety of wireless means. This leads to a number of privacy-related challenges, difficulties, and threats.

### *8.3. Data Exploitation and Representation in STS*

STSs have to deal with a massive amount of data. After extracting the necessary information from the collected data, it becomes necessary to represent the data in a useful way so the traffic management system can get the utmost benefit from it. The system handles a specific type of data to provide some service and does not have to deal with many issues. However, in the case of STS, there are various types of data that need to be integrated into a single system.

### *8.4. Risk Identification*

Systems dealing with various communication and network protocols have high-risk factors. Therefore, any malfunction in the security layers of such a system poses serious threats. In order to operate the system safely, risk identification becomes inevitable. However, due to the massive area coverage of modern megacities,

STS has become huge due to which potential risk identification in every part of the system becomes a serious challenge.

*8.5. Possible Route planning*

The software automatically detects efficient routes for each vehicle and provides dynamic route planning features. Advising and calculating different routes to get off from traffic congestion is the best way to develop the whole traffic efficiency. But the prime challenge is how to execute it at the proper time. Although depending on the central count and entity shows different routes for all transportation to be more effective for its good management along with visual overview, for re-routing and the complication of the algorithm that is utilized in the different route calculations might indicate highly overhead, decreasing its accomplishment depending on the amount of transportation. To solve this problem, there is one solution that enables every vehicle to count its own different routes. But the main difficulty is how to supply a whole overview of traffic. The condition is to enable every transport to count an effective route without overloading the network.

*8.6. Efficiency and Safety Ensuring*

Traffic control is a significant part of current society because it focuses on vehicle supervision to ensure efficiency and safety. It can solve this problem through proper traffic control. Ensuring the confidentiality of information in STS includes government agencies, transit companies, etc. Since the data might be private and can trace information and people and transportation, the main challenge is the actions of environmentally friendly organizations that may include or modify messages which are produced by the services, producing false warning messages from it. There is a new trend of cloud usage that increases complexity in providing computing with STS security of the system and the reason is underlying security concerns with cloud computing have also been included in STS.

*8.7. Risks to Confidentiality and Protection*

Since smart traffic systems depend on ICT and IoT technology and both are based on the Internet, there are privacy and security concerns. The effectiveness of a Smart traffic system depends on uninterrupted connectivity and internet access. As a result, data is more vulnerable and could be stolen. Personal information or even control of the vehicle to some degree may become susceptible to hacking. As IoT connects multiple devices, a virus or malware attack of any kind could prove to be fatal for the whole system and all the devices and vehicles connected [141].

## 9. Discussion

After reviewing and analyzing the research materials selected for this paper and presenting a detailed account of the various aspects of the Smart Traffic system, this section will summarize the findings of this study.

*9.1 Discussion on Approaches in STS*

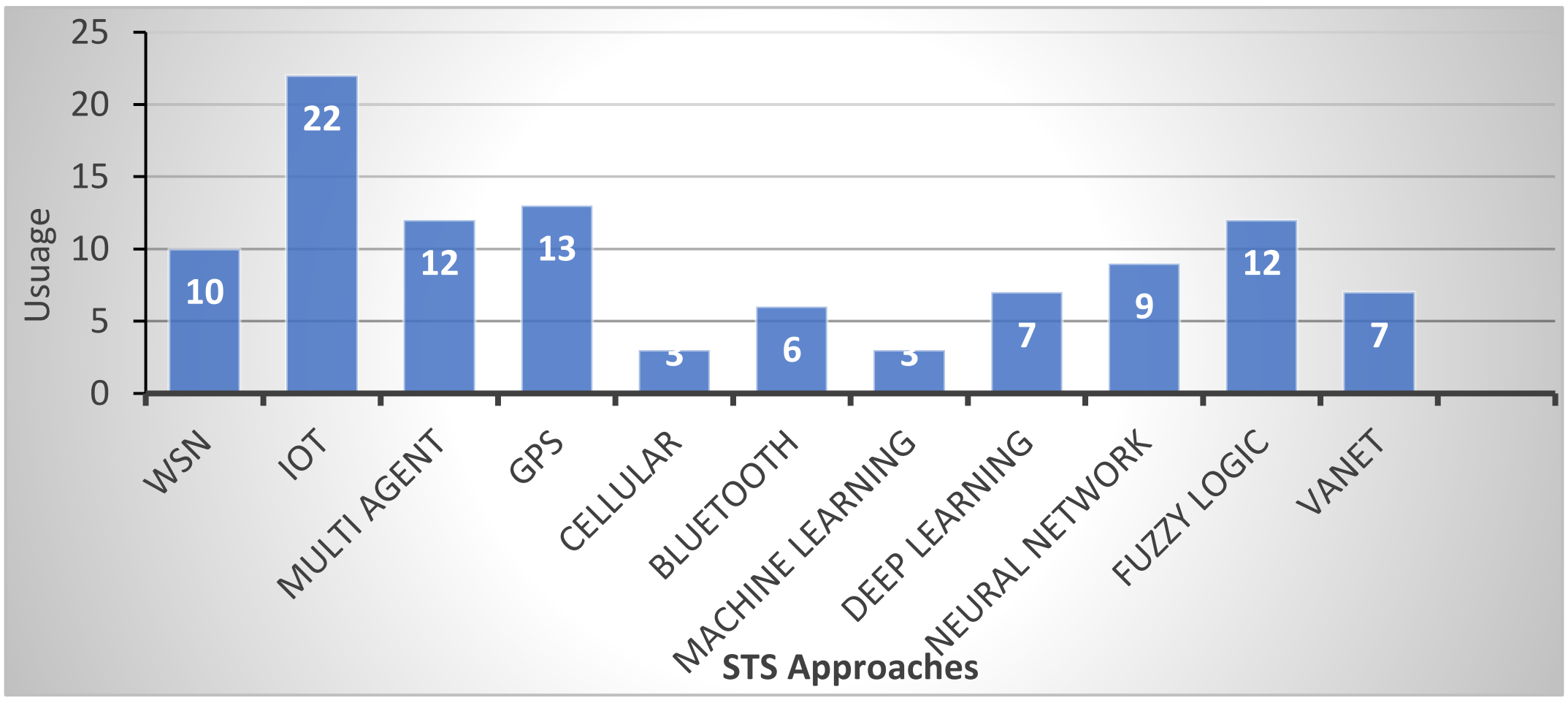

Figure 6. Summarization of Approaches used in STS.

Figure 6 illustrates the technological approaches in STS data gained from table 1. Examining Figure 6, it can easily be identified that IoT is by far the most popular method. This popularity could be attributed to flexibility and reliability as well as the low price of IoT technology [142]. GPS, Multi-Agent, and Fuzzy Logic-based systems are next in line with popularity in STS. Technological advances such as Deep learning, Machine learning, and Neural network methods are also gaining interest among researchers [143]. It should also be mentioned that more than 60% of the research articles that were used in this study used a combination of multiple approaches in their research process.

*9.2 Discussion on Sensors used in STS*

The sensor usage data obtained from table 2 is represented in figure 7, which visually demonstrates the sensor usage trend in STS. It can be seen that the camera module is the most common sensor. This popularity of cameras can be attributed to the fact that a camera can be used in multiple ways in an STS [144-146]. Vehicular sensors and agents are also quite popular in STS. IR and ultrasonic sensors are mostly used for object detection purposes which are gaining popularity due to the increased focus on self-driving technology. Cellular sensors provide access to multiple sensors in a single device thus offering more flexibility and less expense. Also, the usage of magnetometers, RFID, and inductive loop detectors is increasing as vehicle detection and identification is gaining more attention in STS.

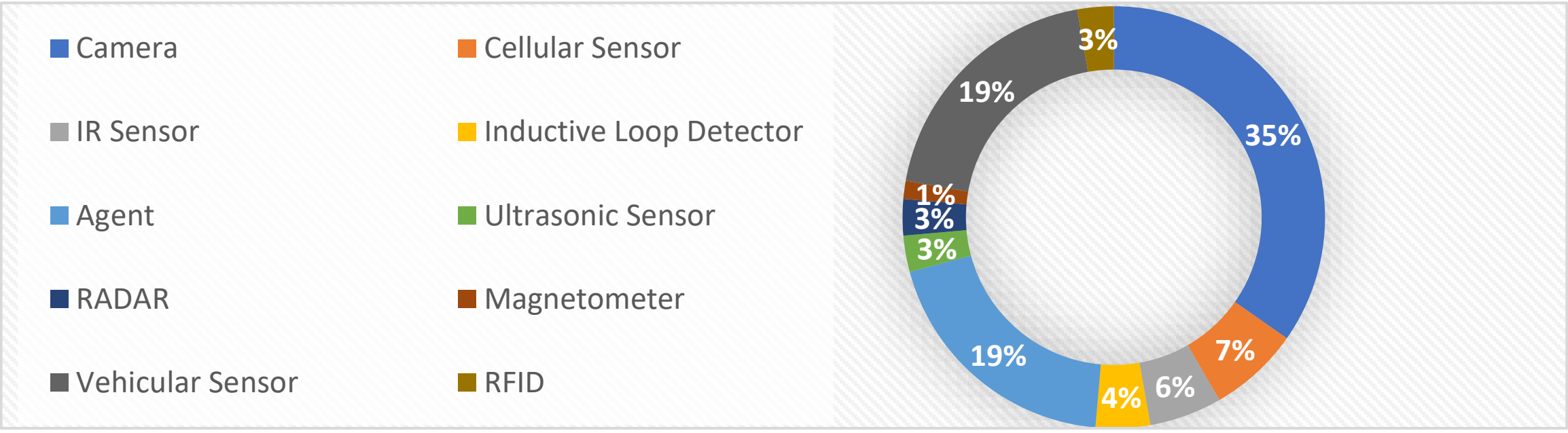

Figure 7. Summarization of Sensors used in STS

*9.3 Discussion on services and suitability of STS*

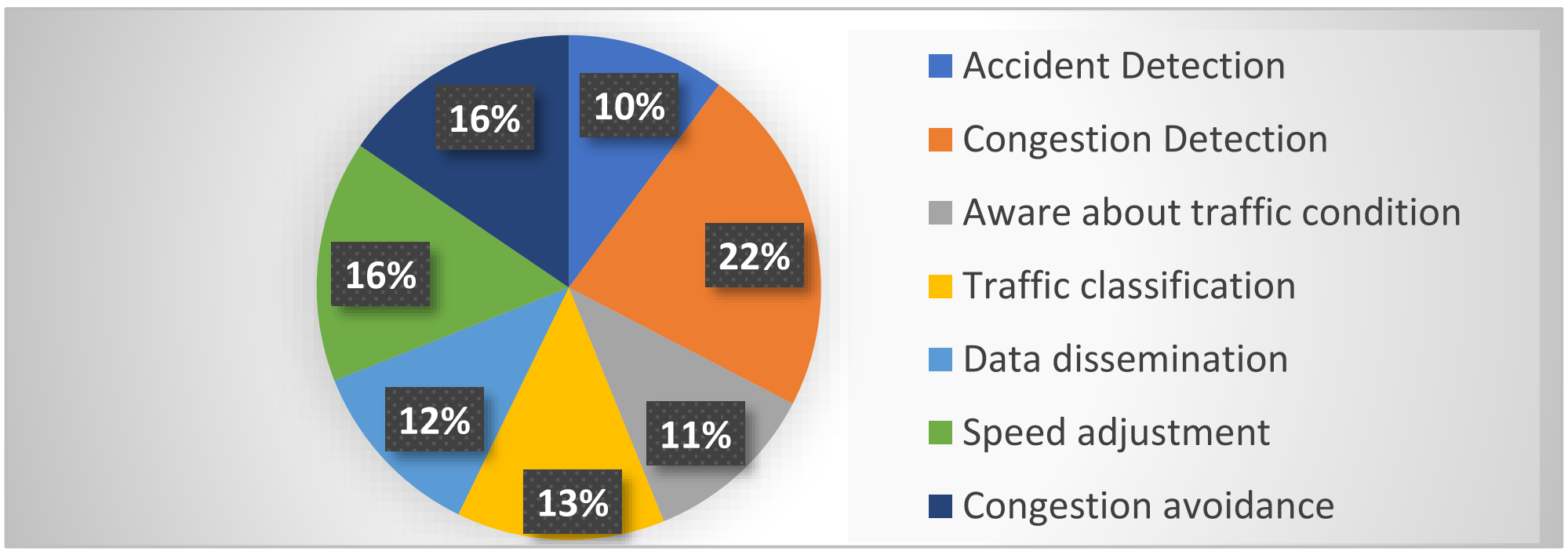

Figure 8. Summarization of STS services & suitability

One of the main sections of this study is the analysis of the services provided by the STS and the adaptability of the systems in different environments. The summarized data about the services of STS is illustrated in figure 8, which highlights the popularity of services. Over the years, services such as accident and congestion detection and traffic classification have become popular. These services help to avoid traffic jams and reach the destination faster and safer. In recent years, computerized automatic services, such as speed adjustment and congestion avoidance have become more popular due to the introduction of autopilot mode in vehicles. To be able to drive, the vehicle needs to know about its surroundings, adjust speed, and avoid any congestion and people. It should be noted that most STS provides multiple services. Also, as discussed in the suitability section, STS for different environments focuses on different problems. Thus, modern STS offers different services depending on the area of implementation.

*9.4. Future Research Scopes in STS*

This study focused on the most recent and relevant research works done on STS. By analyzing the data obtained, the future research trend in this field can be forecasted. Up to this point, most of the research has been focused on solving traditional traffic problems such as traffic congestion, accident avoidance, and traffic signal control for better traffic management [147]. But, with the introduction of autopilot or self-driving cars in recent years, the research has been gradually shifting towards the application of machine learning, artificial intelligence, and deep learning in order to develop more robust and intelligent self-driving systems [148-152]. By automating the driving process, it is now possible to set up a totally computerized STS that will control both the vehicles and road signals in the optimum possible way to decrease traffic congestion. From the discussion, it is clear that IoT, with the help of GPS and other technological methods, will prove to be the backbone of STS in future research [153-154]. By creating a network of all the vehicles on the road, it will be possible to avoid accidents and thus create a safety net for all the vehicles [155]. The usage of sensors, especially cameras, IR, magnetometers, RADAR, etc. can help with detecting and identifying objects and vehicles. Also, the information about the surroundings of a vehicle will increase. These types of sensors will prove to be essential in future research focuses due to increased attention toward AI. Future research endeavors on STS will continue to shift towards creating a totally autonomous driving system [156-160]. Inter-vehicles communication with the help of IoT and the application of AI and machine learning will be the key components of the future STS. On the other hand, due to increased dependency on networking and internet-based systems, the risk of data theft and online security will become a serious problem [161-165]. Hacking into self-driving vehicles and taking control of the system by miscreants could very well pose serious threats [166]. Additionally, AI, machine learning, and deep learning systems might not be able to come up with accurate predictions for the best route plan to avoid traffic congestion due to insufficient data and faulty algorithms, which may end up causing serious traffic problems if there is no human involvement [167-173]. These are some of the problems that will need to be solved before the full-scale implementation of STS.

## 10. Conclusion

The massive amount of traffic in megacities and highways poses a great challenge to the traffic management method due to which various STSs have evolved. To provide an overall overview of the STS, a comprehensive review work based on information extraction from existing research papers has been provided in this research. The STS is classified as per various aspects. First, SPS classification based on different approaches has been provided. Then, a detailed analysis of the various sensors utilized in STS has been provided. Next, the communication and networking protocols utilized in STS are highlighted along with their specifications. After that, different services provided by present day STS are featured. Finally, challenges associated with STSs are identified so that future research work can emphasize them. The comprehensive review conducted in this research is expected to provide a bird’s eye view of STS to the researchers, industry persons, system designers and policymakers.

## Conflict of Interest

The authors declare no potential conflict of interest.

## Authorship Contribution Statement

**Arindom Chakraborty:** Conceptualization, Methodology, Formal Analysis, Investigation, Writing-Original Draft, **Mehedi Hasan:** Conceptualization, Methodology, Investigation, Writing-Original Draft, **Amzad Hossain:** Methodology, Investigation, Writing-Review, Editing & Supervising, **Meratun Junnut Anee:** Methodology, Investigation, Writing-Review & Editing.